\pdfoutput=1
\documentclass[%
prx,twocolumn,
notitlepage,
superscriptaddress,
amsmath,
amssymb,
aps,
prx,
floatfix,
longbibliography
]{revtex4-1}
\usepackage[T1]{fontenc}

\usepackage{color}
\usepackage{amsmath}
\usepackage{amssymb}
\usepackage{bm}
\usepackage{graphicx}
\usepackage[usenames,dvipsnames]{xcolor}
\usepackage[%
  colorlinks=true,
  urlcolor=blue,
  linkcolor=blue,
  citecolor=blue
]{hyperref}

\usepackage{array}
\definecolor{bleu}{rgb}{0.16,0.2.5,0.36}
\definecolor{darkGreen}{RGB}{4,161,85}

\newcommand{\blue}[1]{{\color{blue} #1}}

\newcommand{\EDF}{Extended Data Fig.}

\begin{document}
\pdfoutput=1

\title{Observation of strain-rate softening behavior in jammed granular media}

\author{Mingchao Liu}
\email{m.liu.2@bham.ac.uk (M.L.)}
\affiliation{These two authors contributed equally.}
\affiliation{School of Mechanical and Aerospace Engineering, Nanyang Technological University, Singapore, 639798, Republic of Singapore}
\affiliation{Department of Mechanical Engineering, University of Birmingham, Birmingham B15 2TT, UK}
\author{Weining Mao}
\affiliation{These two authors contributed equally.}
\affiliation{School of Mechanical and Aerospace Engineering, Nanyang Technological University, Singapore, 639798, Republic of Singapore}
\author{Yiqiu Zhao}
\affiliation{Department of Physics, The Hong Kong University of Science and Technology, Hong Kong SAR, China}
\author{Qin Xu}
\affiliation{Department of Physics, The Hong Kong University of Science and Technology, Hong Kong SAR, China}
\author{Yixiang Gan}
\affiliation{School of Civil Engineering, The University of Sydney, Sydney, NSW, Australia}
\author{Yifan Wang}
\email{yifan.wang@ntu.edu.sg (Y.W.)}
\affiliation{School of Mechanical and Aerospace Engineering, Nanyang Technological University, Singapore, 639798, Republic of Singapore}
\author{K Jimmy Hsia}
\email{kjhsia@ntu.edu.sg (K.J.H.)}
\affiliation{School of Mechanical and Aerospace Engineering, Nanyang Technological University, Singapore, 639798, Republic of Singapore}
\affiliation{School of Chemistry, Chemical Engineering and Biotechnology, Nanyang Technological University, Singapore, 639798, Republic of Singapore}

\date{April 15, 2024}
\maketitle

{\bf 
The strain-rate sensitivity of confined granular materials has been widely explored,  with most findings exhibiting rate-strengthening behaviors.
This study, however, reveals a distinct rate-softening behavior across a certain strain rate range based on triaxial tests on particle clusters of various materials with different surface properties, particle sizes, shapes, and stiffness. This softening effect is especially pronounced in the case of common rice particles.
By examining the behavior of rice particles under different confining pressure and surface conditions, and directly measuring the frictional coefficient across various loading rates, we find that the reduction in surface frictional coefficient with the increasing strain rate predominantly contributes to this rate-softening behavior. This conclusion is validated by results from Finite Element Method (FEM) simulations. Additionally, we identify confining pressure as a critical factor regulating the normal stress between particles, and thereby enhancing frictional behavior. 
Rheometer tests reveal that the shear modulus exhibits a similar rate-softening trend.
This study of rate-softening behavior in granular materials enhances our understanding of the mechanisms during their deformation under confining pressure. It also suggests that local inter-particle tribology significantly impacts overall granular behavior.}
\\

The strain-rate dependence in mechanical responses of solid materials is ubiquitous, with most exhibiting strain-rate strengthening behavior; that is, their stress increases with increasing strain rate. The jamming of granular materials, which transition from a fluid-like to a solid-like state under sufficient confining pressure \cite{RN-45-majmudar2007jamming,xing2024origin}, also typically exhibits strain-rate strengthening. This strain-rate strengthening behavior has been experimentally observed in many granular systems, including sand \cite{RN4-Duttine2009,RN5-Suescun-Florez2017,RN6-Mukherjee2020}, soil \cite{RN8-Enomoto2016,RN7-Soni2022}, photoelastic polymers \cite{RN10-hartley2003logarithmic,RN11-Behringer2008}, and coffee beans \cite{RN12-wangY2019}, etc. In these materials, the corresponding deviatoric stress is observed to increase with increasing applied strain rate during triaxial loading.
In contrast, snow and puffed rice are two granular materials reported to exhibit recognizable strain-rate softening behavior \cite{RN19-Valdes2017,RN20-Narita2017,RN21-Barraclough2016}. Previous researches suggest that particle crushing may be the dominant factor in this softening behavior in these brittle materials, which cannot be generalized to wider granular systems.


In this paper, we present our observations of a remarkable strain-rate softening behavior in solid non-brittle granular materials, which has not been reported in the past. Intriguingly, this behavior is observed in everyday materials like common rice particles. We conduct comprehensive experiments and simulations to explore the mechanisms governing the observed rate-softening behavior, aiming to identify the primary factors contributing to this phenomenon. Our research enhances our understanding of deformation instability in granular materials, which may provide insights into natural disasters like earthquakes \cite{RN34-reches2010fault,RN35-rice1985constitutive} and avalanches \cite{RN-51-daerr1999two}. Moreover, understanding rate-softening may benefit practical applications, such as in the development of impact protection systems or in optimizing processes that involve granular flow, like pouring, by minimizing resistance.

\begin{figure}[h!]
\centering
\includegraphics[width=1.0\linewidth]{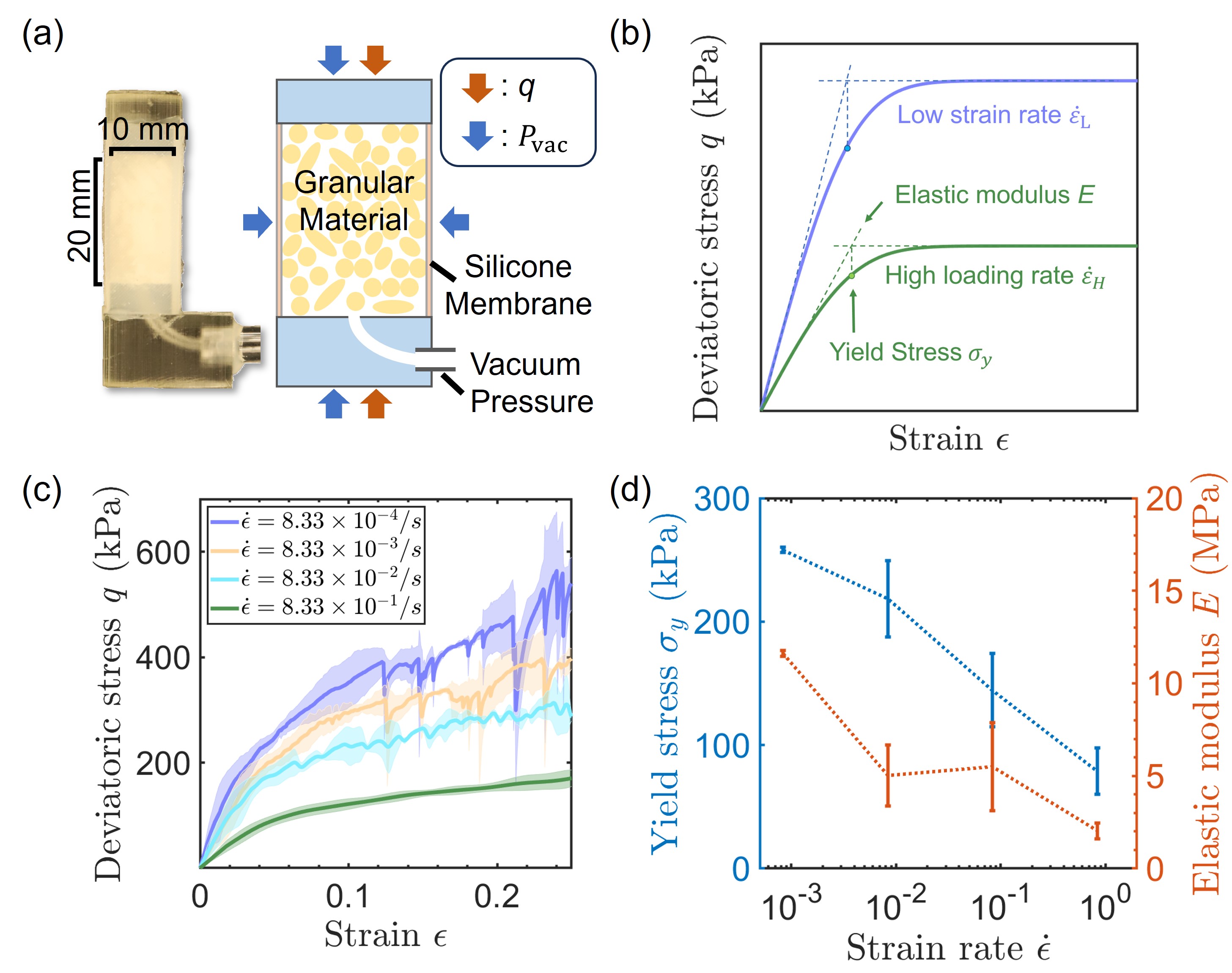}
\caption{\textbf{Experimental setup and the strain-rate softening behavior.} 
\textbf{(a)} A cuboid column sample with dimensions 10 mm $\times$ 10 mm $\times$ 20 mm packing of rice particles in a silicone membrane and evacuated to a vacuum pressure $P_{vac}$ before a load with constant strain rate is applied. The deviatoric stress $q$ is measured.
\textbf{(b)} A schematic of the typical curve of the rate-softening behavior observed during experiments. To evaluate the intensity of this behavior, a yield point is picked as shown in the figure.
\textbf{(c)} Plot for deviatoric stress $q$ vs strain $\epsilon$ for the specimen filled with rice particles under strain rate ranging from $8.33\times{}10^{-4}$/s to $8.33\times{}10^{-1}$/s. The vacuum pressure is 60 kPa.
\textbf{(d)} Both the yield stress $\sigma_y$ (left axis in blue) and the elastic modulus $E$ (right axis in red) of the rice particles decrease gradually with the increase of strain rate in the log scale.}
\label{fig:1}
\end{figure}

\begin{figure*}[t!]
\centering
\includegraphics[width=0.75\linewidth]{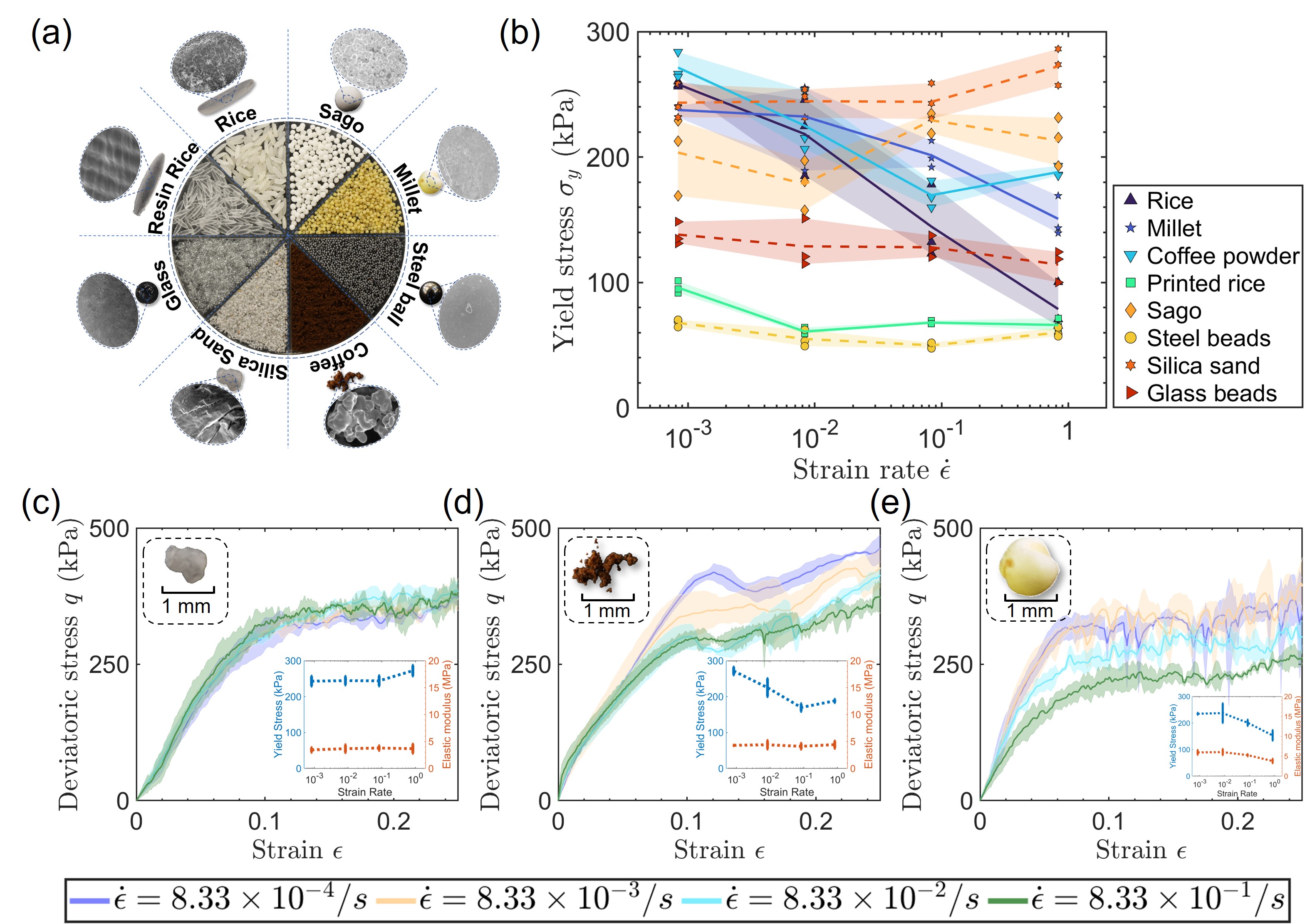}
\caption{\textbf{Strain-rate dependent behaviors of different granular materials.} 
\textbf{(a)} Eight granular materials with different particle shape, particle size, and surface morphology are tested. \textbf{(b)} Plots for the yield stress $\sigma_y$ vs strain rate $\dot{\epsilon}$ for all eight materials. The materials with obvious rate-softening behavior are marked with solid lines, while the materials with rate-independent or rate-strengthening behavior are marked with dash lines. The deviatoric stress $q$ vs strain $\epsilon$ relation for \textbf{(c)} silica sand which shows slightly rate-strengthening behavior in this strain rate range; for \textbf{(d)} coffee powder and \textbf{(e)} millet which shows rate-softening behavior.
}
\label{fig:2}
\end{figure*}

As depicted in Fig. \ref{fig:1}(a), we perform triaxial compression tests on granular materials confined in a silicone membrane column under vacuum pressure $P_{vac}$, subjected to different strain rates, $\dot{\epsilon}$. Detailed experimental procedures and material parameters are available in {\color{blue}Supplementary Information Section A}. 
Figure \ref{fig:1}(b) presents a schematic diagram that illustrates the rate-softening behavior. In this test, the apparent Young's modulus, $E$, and the yield stress, $\sigma_y$, are measured. 
Specifically, $\sigma_y$ is determined by locating the intersect of the extended initial elastic portion and the plastic portion of the stress-strain curve. This yield point represents the point at which the granular system begins to have permanent deformation in the jammed state \cite{RN-47-behringer2018physics}.
The intensity of the rate-softening is quantified by the extent of the decrease in yield stress $\sigma_y$ as the strain rate $\dot{\epsilon}$ increases. 
Take a sample composed of rice particles as an example (Fig. \ref{fig:1}(a)), a pronounced rate-softening behavior is observed, as shown in Figs. \ref{fig:1}(c) and (d). It is evident that both elastic modulus $E$ and yield stress $\sigma_y$ decrease significantly with an increase in strain rate $\dot{\epsilon}$, dropping to approximately one-third of their original values when $\dot{\epsilon}$ varies by three orders of magnitude from $8.33\times10^{-4}$/s to $8.33\times10^{-1}$/s.

To establish generality, we expand our investigation to encompass a total of eight granular materials, as summarized in Fig. \ref{fig:2}(a). Each material is selected for its unique particle shape, size, stiffness, and surface properties. The relationship between yield stress $\sigma_y$ and strain rate $\dot{\epsilon}$ for all eight materials is depicted in Fig. \ref{fig:2}(b). Within the specified strain rate range ($\dot{\epsilon} \sim 8.33\times10^{-4}-8.33\times10^{-1}$/s), silicon sand, sago, and steel balls show behaviors that are either rate-independent or slightly rate-strengthening. In contrast, rice, millet, coffee powder, glass beads, and the resin-printed rice-shaped particle (which is referred to as printed rice) demonstrate notable rate-softening behavior. In particular, the rate-softening of rice is most profound among these materials. 
We further present the stress-strain relations of silicon sand (Fig. \ref{fig:2}(c)), coffee powder (Fig. \ref{fig:2}(d)), and millet (Fig. \ref{fig:2}(e)) as examples, which represent the rate-independence or rate-softening samples. Additionally, stick and slip behavior can be observed in both rice and millet at three low strain rates as seen in the serrated stress-strain curve. Notably, the stress drops due to these stick-slip behaviors do not go below the stress curve with the highest strain rate ($\dot{\epsilon} \sim 8.33\times10^{-1}$/s). During the  stick-slip process, local instability may cause the local strain rate to increase significantly and cause a drop in the stress. The results for the other four materials (i.e., printed rice, glass beads, steel beads, and sago), as well as the experiments that investigate the boundary effect, are shown in \blue{Supplementary Information Section B.}

To understand the underlying mechanisms, we focus on the rice particle as it possesses the strongest rate-softening behavior in the specified strain rate range. Fig. \ref{fig:1}(c) shows that  rice samples exhibit a stick-slip behavior under low strain rate, which is a typical instability behavior observed in frictional granular materials \cite{RN16-Chambon2002,RN17-Ozbay2016}. On the other hand, the stress-strain curves are rather smooth under high strain rate, similar to the behaviors in frictionless granular materials \cite{RN-48-maloney2008evolution,RN-47-behringer2018physics}. 

\begin{figure*}[t!]
\centering
\includegraphics[width = 0.75\linewidth]{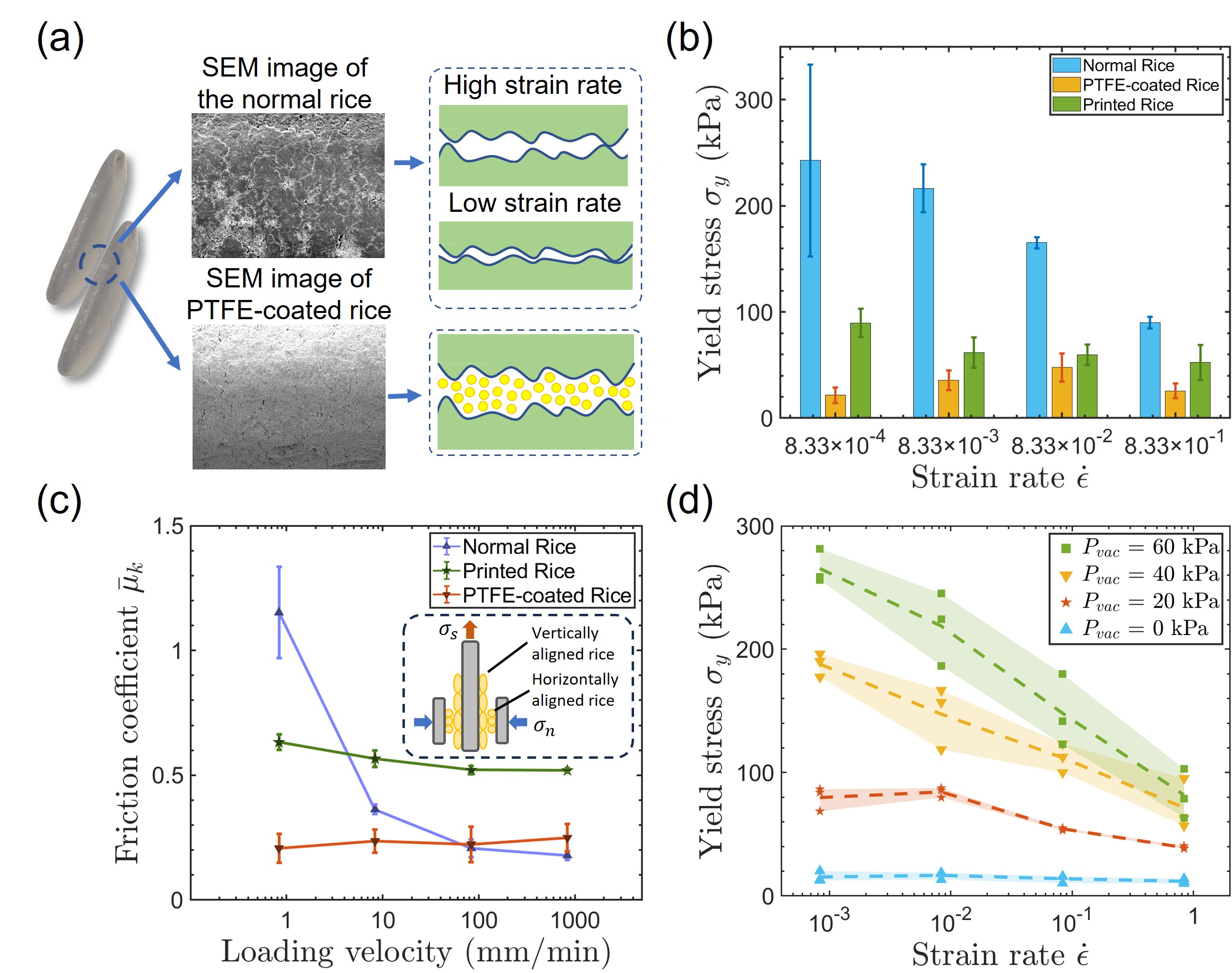}
\caption{\textbf{Main factors that may cause this rate-dependent behavior.} \textbf{(a)} the SEM image and schematic diagram of the rice surface with and without PTFE. The coated PTFE prevents the creep behavior which reduce the rate-dependency of the surface friction. 
\textbf{(b)} the yield stress $\sigma_y$ vs strain rate $\dot{\epsilon}$ for normal rice, rice coated with PTFE, and printed rice manufactured by resin. 
\textbf{(c)} The average friction coefficient $\bar{\mu}_k$ varies with loading velocity $v$. The granular material is aligned and glued to a slab to obtain an almost flat friction surface. The middle slab is then sheared upward at a constant strain rate while a 3N normal force is applied on both sides of the other two slabs. 
\textbf{(d)} Plot of the yield stress $\sigma_y$ vs strain rate $\dot{\epsilon}$ during loading under applied vacuum pressure $P_{vac}$ of 60 kPa, 40 kPa, 20 kPa, and 0 kPa.
}
\label{fig:3}
\end{figure*}

To  explore the effect of surface friction, we coated the rice particles with Polytetrafluoroethylene (PTFE,  which acts as a solid lubricant) and conducted the triaxial compression test within the same strain rate range. As shown in Fig. \ref{fig:3}(a), the surface morphology of PTFE-coated rice is much smoother than the uncoated ones and, consequently, should exhibit reduced interfacial friction. The dependencies of yield stress $\sigma_y$ on the strain rate $\dot{\epsilon}$ during dynamic compression are presented in Fig. \ref{fig:3}(b). Compared to the strong rate-softening behavior of natural rice, the PTFE-coated rice shows a nearly constant yield stress. Considering the identical properties of the rice particle and the loading condition in the two experiments, we conclude that surface friction is a dominant factor for this rate-softening behavior.

To quantitatively evaluate the surface properties of rice particles, an experiment is designed to measure the frictional coefficient. As shown in the inset of Fig. \ref{fig:3}(c), we clip two panels with rice glued on one side to a middle panel that has rice glued on both sides. A normal confining load is applied in the horizontal direction of the side panels while the middle panel is pulled up in the vertical direction with four different loading rates. The normal pressure is measured during the loading process. The detailed setup for this experiment is shown in {\color{blue}Supplementary Information Section C}. The equivalent mean dynamic friction coefficient $\mu$ is calculated by
\begin{align}
    \mu=\frac{\sum_{i=1}^{n}{\frac{F_i}{N_i}\Delta t_i}}{\sum_{i=1}^{n}{\Delta t_i}},
\end{align}
where $F_i$ and $N_i$ are the vertical force and normal force measured at the time $t_i$ respectively. $\Delta t_i$ is the time interval between the data points at time $t_i$ and $t_{i-1}$. Considering the loading rate is controlled as a finite constant value in each test, we consider all the friction coefficients as dynamic friction coefficients.

As shown in Fig. \ref{fig:3}(c), the friction coefficient of original rice particles significantly drops from 1.18 to 0.19 as the loading velocity increases from 1mm/min to 1000 mm/min. In contrast, with the same loading velocity variation, the coefficient for PTFE-coated rice remains nearly constant. We also measured the friction coefficient of printed rice. The dependence of this coefficient on the loading rate falls between that of normal rice and PTFE-coated rice, aligning with the observed moderate strain-rate softening of yield stress (Fig. \ref{fig:2}(c)).

For the frictional behavior, we consider the Rate and State Friction (RSF) law with a rate-dependent factor and a time-dependent healing factor  \cite{RN24-scholz1998earthquakes,RN25-Dieterich1979}, i.e., 
\begin{align}
\mu=\mu_0+A\times\ ln(\frac{v}{v_0})+B\times\ ln(\frac{\phi}{\phi_0}),
\end{align}
where $\mu$ is the friction coefficient, $v$ is the loading velocity, $v_0$ is a reference velocity, $\mu_0$ is the static friction coefficient when $v = v_0$, $A$ and $B$ are material parameters, $\phi$ is the state variable representing the aging of the contact surface, which describes the relationship between friction and surface contacting time, defined as
\begin{align}
\dot{\phi}=1-\frac{v\phi}{D_0},
\end{align}
where $D_0$ is the distance of sliding when the new contact is established under sliding speed $v_0$. In our case, we apply a stationary load at a constant strain rate which gives $\phi=D_0/v$. Therefore, the friction coefficient can be expressed as
\begin{align}
\mu=\mu_0'+(A-B) \times ln(v).
\end{align}

Note that the value of $(A-B)$ serves as a quantitative measure of the rate-dependence of the friction coefficient \cite{RN24-scholz1998earthquakes,RN25-Dieterich1979}.
From Fig. \ref{fig:3}(c), we measure the value of $(A-B)$ as $-0.307$ for original rice, $-0.038$ for printed rice, and $0.011$ for PTFE-coated rice. Note that the response of these particles does not exhibit a perfectly linear logarithmic relationship as reported for glass or rocks \cite{RN26-Murphy2019}. This deviation may be attributed to their elongated shape and the surfaces not being perfectly flat during sliding. Additionally, the PTFE-coated rice particle has a significantly larger elastic modulus compared to the natural rice, which may prevent the contact of the original rice surface so that it gives a nearly zero value of $(A-B)$.

Comparing the value of $(A-B)$ from Fig. \ref{fig:3}(c) with the rate-softening behavior of the yield stress in Fig. \ref{fig:3}(b), we find that their trends are consistent with each other, i.e., a larger magnitude of $\left| (A-B) \right|$ corresponds to a higher drop of yield stress with strain rate for rice.
Additionally, the results suggest that surfaces with negative $(A-B)$ values will lead to oscillatory nature of the stick-slip bifurcation \cite{RN-44baumberger1999physical},  resulting in the strong stick-slip bulk behavior of rate-softening materials.

At low loading rates, the high friction coefficient between particles helps to build up strong force chains which result in high initial elastic modulus and yield stress. When the loading rate increases, the decrease of the friction between particles weakens the formation of force chains and causes the reduction in the initial elastic modulus. Besides, previous experimental and numerical studies on frictional jamming reported that changes of the friction coefficient would change the magnitude of the yield line between the jammed state and the flow state \cite{RN27-Song2008,RN28-Ciamarra2011,RN29-Bi2011}. Therefore, when the strain rate acting on the rice particle cluster increases, the drop of friction coefficient shifts the stress level at yielding, and causes the system to transition from a jammed state to a flow state under a smaller shear stress, giving rise to the rate-softening behavior. Meanwhile, the high friction coefficient under low loading rate causes the system to have a hyper-static frictional jamming behavior which causes the stick-slip behavior at low loading rate because of the local collapse of the force chain caused by the instability induced by the rate-weakening friction \cite{RN30-Leeman2015}. Since this drop of the friction has a limit and cannot decrease below the value corresponding to a high strain rate, the  stress level at $8.33\times{10}^{-4}$/s never drops below the stress level at $8.33\times{10}^{-1}$/s. In contrast, under high loading rates, the friction coefficient of the rice particles is considerably lower, which causes the system to behave as an isostatic frictionless jamming state, and thus a more stable shear response.

We have now shown that the rate-dependent friction coefficient is an important factor governing the rate-softening behavior. Since the frictional behavior of granular material is coupled with the confining pressure\cite{RN27-Song2008,RN28-Ciamarra2011}, it is expected that the confining pressure would affect the rate-softening behaviors.
We conducted triaxial compression experiments on samples filled with common rice under four different confining pressure levels. The dependencies of yield stress, $\sigma_y$, on strain rate, $\dot{\epsilon}$, under different confining pressure are summarized in Fig. \ref{fig:3}(d). It is observed that the rate-softening behavior is amplified by higher confining pressure. Specifically, as the confining pressure is reduced from 60 kPa to 0 kPa, the degree of rate-softening for rice decreases significantly, eventually becoming rate-independent at zero confining pressure.
This is because higher confining pressure results in higher friction force between particles, leading to a more significant rate-dependent behavior.

Additionally, we also investigate the effect of particle shape, which affects the area and shape of the contact surface, on the rate-softening behavior, and found that the particle shape has limited influence on the rate dependency, as can be seen in \blue{Supplementary Information Section D.}

\begin{figure}[h!]
\centering
\includegraphics[width=1.0\linewidth]{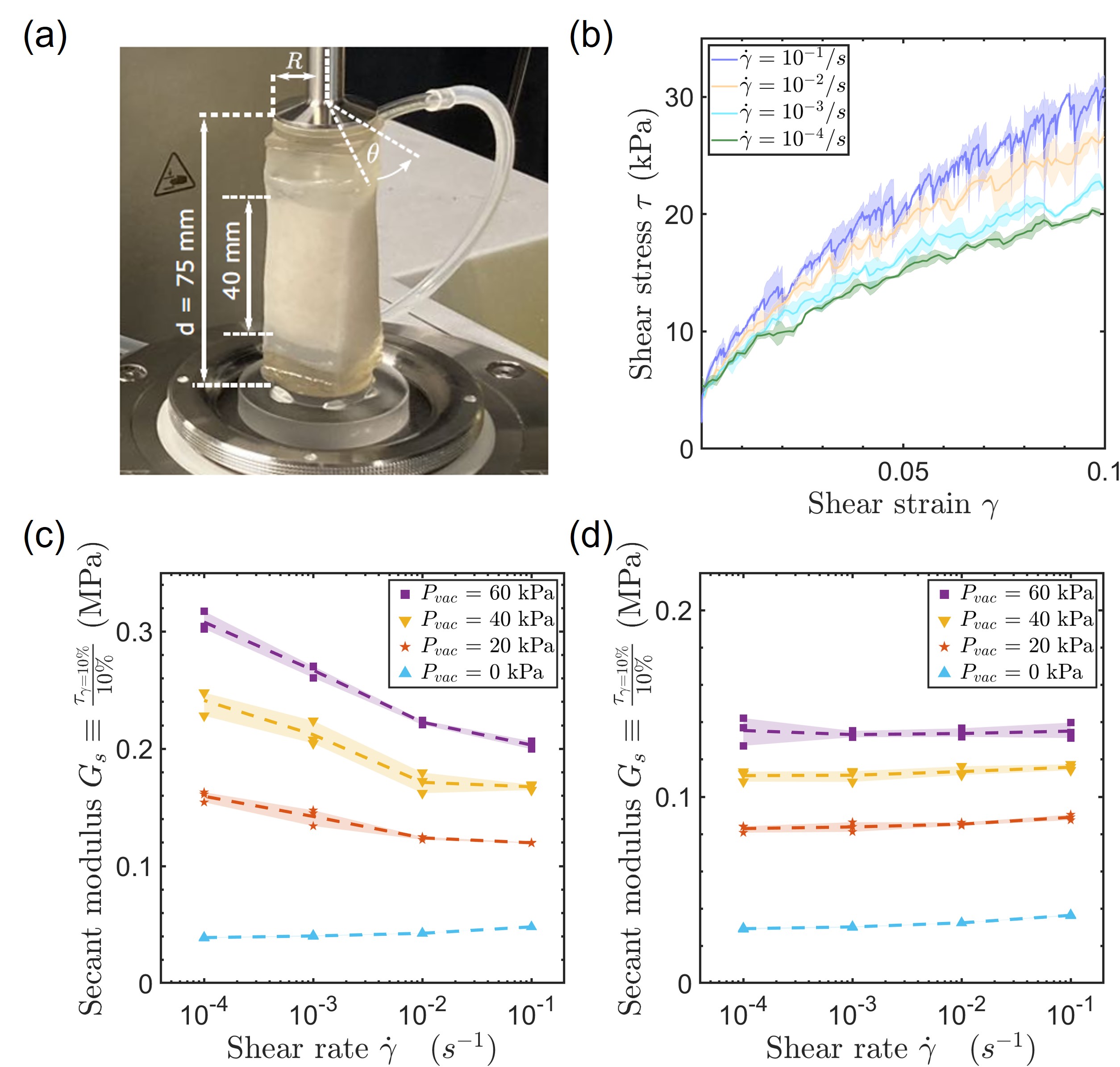}
\caption{\textbf{Rate-softening shear behavior observed in rheometer test.}
\textbf{(a)} The rheometer is used to measure the rate-softening shear behavior of the rice particles. A sample with dimensions 20 mm $\times$ 20 mm $\times$ 40 mm packing of normal rice particles sample is glued to the rheometer and a shear strain was applied with constant shear rate.
\textbf{(b)} The sample with normal rice was subjected to 4 different shear rates $10^{-1}$/s, $10^{-2}$/s, $10^{-3}$/s, and $10^{-4}$/s. 
\textbf{(c)} The secant shear modulus at $\gamma=10\%$ was calculated to evaluate the rate-dependent shear behavior. The figure plots the secant shear modulus $G_s$ vs shear strain rate $\dot{\gamma}$ under the vacuum pressure $P_{vac}$ of 60 kPa, 40 kPa, 20 kPa, and 0 kPa for normal rice.
\textbf{(d)} The plots of secant shear modulus $G_s$ vs shear strain rate $\dot{\gamma}$  under the vacuum pressure $P_{vac}$ of 60 kPa, 40 kPa, 20 kPa, and 0 kPa for PTFE-coated rice.
}
\label{fig:4}
\end{figure}

We further carry out the rheometer test of large samples of rice particles \cite{RN-50-de2017rheology},
%
 as shown in Fig. \ref{fig:4}(a). To measure the shear stress, $\tau$, samples subjected to a confining pressure, $P_{\text{vac}}$, ranging from 0 kPa to 60 kPa, are rotated at four different shear rates, $\dot{\gamma}$, from $10^{-1}$/s to $10^{-4}$/s. The detailed experimental method is described in {\color{blue}Supplementary Information Section E}.
Figure \ref{fig:4}(b) shows the shear stress $\tau$ versus shear strain $\gamma$ for different shear rates $\dot{\gamma}$ under a confining pressure of $P_\text{vac} = 60$ kPa. Similar to observations in the compression experiments, there is a clear rate-softening trend, with shear stress drops of  larger magnitude at lower frequency due to stick-slip.

To quantify the rate-dependence, the secant modulus at $\gamma_m = 10\%$ is examined, defined as the ratio of $\tau$ to $\gamma$ at this point. As depicted in Fig. \ref{fig:4}(c), the rate-softening behavior under shear strain becomes more pronounced at higher confining pressures, while the sample exhibits rate-independent behavior in the absence of confining pressure (i.e., $\sigma_\text{vac} = 0$ kPa). When coated with PTFE, the rice particles exhibit rate-independent behavior regardless of the level of the confining pressure, as illustrated in Fig. \ref{fig:4}(d). The results from the rheometer test are consistent with those from the compression test, supporting the conclusion that the rate-dependent friction coefficient of the particle surface is the primary cause of the observed rate-softening behavior in the jammed granular media.

\begin{figure}[h!]
\centering
\includegraphics[width=1.0\linewidth]{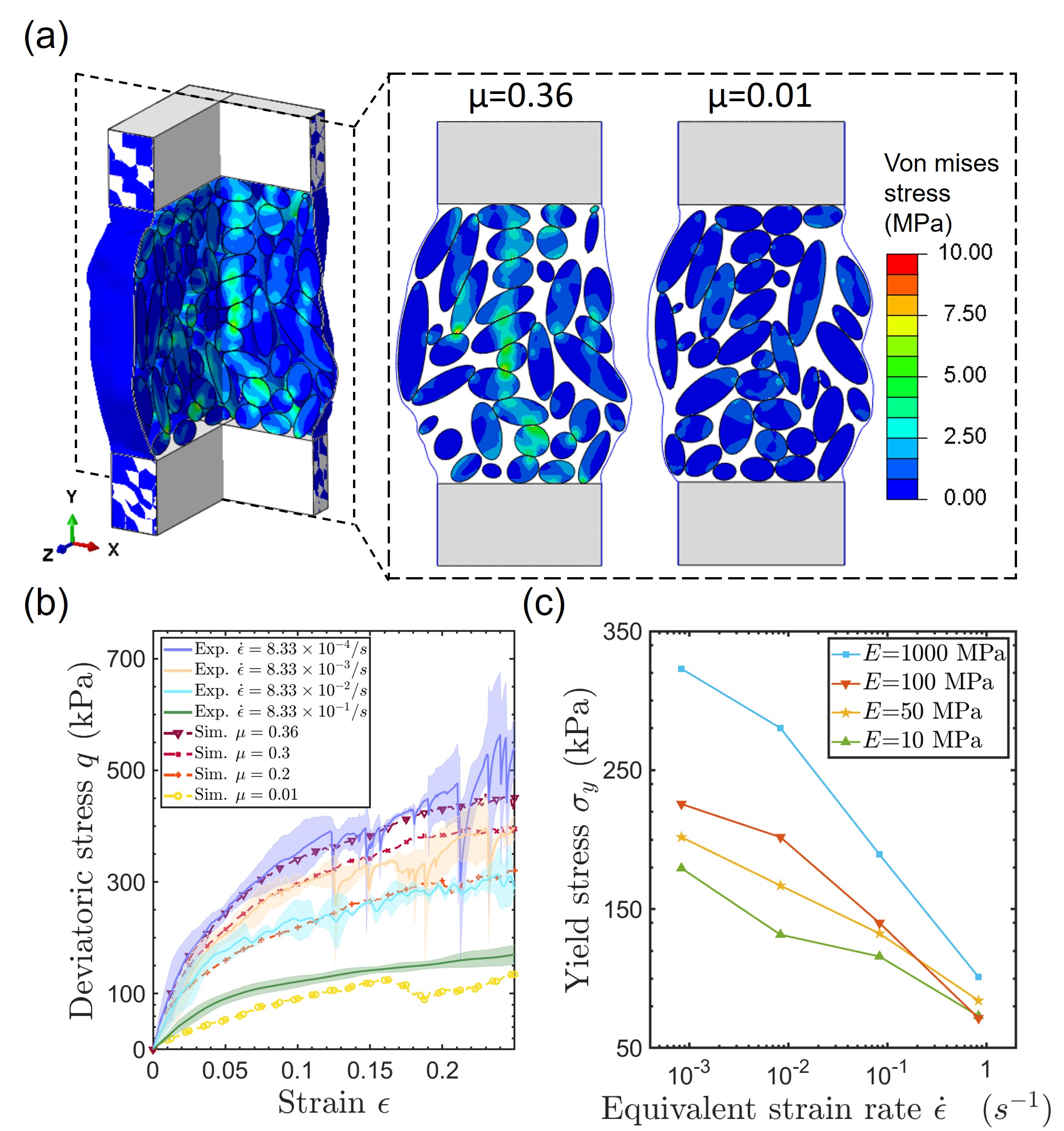}
\caption{\textbf{FEM simulation of the sample filled with rice particles.} 
\textbf{(a)} Cross-section plot of the Von Mises stress of the FEM simulation at $\epsilon=0.167$ for (left) $\mu=0.36$ and (right) $\mu=0.01$ under a confining pressure of 60 kPa. 
\textbf{(b)} Comparing the experiment data (Exp.) with the FEM simulation (Sim.) of the vacuumed rice particle sample in which the decrease of the friction coefficient is used to represent the increase of loading strain rate. 
\textbf{(c)} By representing the equivalent strain rate by the friction coefficient in FEM simulation, we plot the drop of yield stress under different particle elastic modulus $E$.  
}
\label{fig:5}
\end{figure}

To further investigate the effect of the surface friction coefficient on the collective strain-rate softening behavior, we carry out Finite Element Method (FEM) simulations of the compression loading of the granular rice sample under confining pressure, as demonstrated in Fig. \ref{fig:5}(a). The increase of the strain rate is represented by the decrease of the frictional coefficient. The detailed simulation method is reported in {\color{blue}Supplementary Information Section F}. 
The FEM simulation results with $\mu$ decreasing from 0.36 to 0.01 in Fig. \ref{fig:5}(b) show a very similar trend to the experimental results with strain rate $\dot{\epsilon}$ increases from $8.33\times{10}^{-4}$/s to $8.33\times{10}^{-1}$/s. Meanwhile, Fig. \ref{fig:5}(a) depicts the stress distributions inside a sample with $\mu=0.36$ and $\mu=0.01$, demonstrating that the  rice particles with higher friction coefficients form  stronger force chains. These results provide further evidence of our argument that the reduction of friction coefficient when increasing the loading rate will cause the weakening of the force chains in the particle cluster,  leading to rate-softening behavior. 
It's important to note that the values of the friction coefficient used in the FEM simulations significantly deviate from the measure ones in Fig. \ref{fig:3}(c). This discrepancy is likely attributed to the fact that the local relative velocity between the surfaces of two particles generally does not correspond directly to the applied loading rate, as shown in Fig. \ref{SI7_FEM}(f) in {\color{blue}Supplementary Information Section F}.

Consistent with the experimental measurements, the FEM simulation results show that high friction coefficient helps build up strong force chains, as shown in Fig. \ref{fig:5}(a), giving rise  to both high initial elastic modulus and high stresses beyond yielding. Additionally, we investigate the influence of the elastic modulus of particles on the rate-dependent behavior with FEM simulations. The friction coefficient values employed in Fig. \ref{fig:5}(b) are used to represent the corresponding strain rates. As illustrated in Fig. \ref{fig:5}(c), increasing the elastic modulus of granular particles while keeping the surface and geometry unchanged also leads to a more pronounced rate-softening behavior. However, it is important to note that an increase in elastic modulus may also affect the aging of the contact surface, which is not accounted for in this study. The contribution of the elastic modulus of particles to enhanced rate-softening behavior in granular media needs to be confirmed by further experimental validation.

In conclusion,  we report the observation of pronounced strain-rate softening behavior in granular materials composed of various types of particles, including common materials such as rice.
Through experiments and FEM simulations, we have provided evidence  that this strain-rate softening behavior is primarily due to the rate-dependence of the friction coefficient, originating from the surface aging of the particles. The local aging effect between particle surfaces builds up a stronger force chain network inside the granular cluster so that the lower loading rate gives a higher stress response. Additionally, higher confining pressures lead to more pronounced rate-softening behavior, while the shape of the particles plays a negligible role. 
Our findings contribute to a better understanding of the rate-dependent mechanical responses of jammed granular systems. It also shows that the local rate-dependent tribology between particles plays a vital role in the rate-dependent behavior of granular material, a mechanism rarely mentioned in the research literature.
%
Furthermore, these insights enable the design of intelligent granular protection devices that, e.g., can automatically adapt their responses based on impact velocity. 

\emph{Acknowledgments.} 
We thank Itai Einav, Daniel Bonn, Edan Lerner, Bart Weber, Corentin Coulais, Dominic Vella and Si Suo for fruitful and illuminating discussions. We acknowledge financial supports from Nanyang Technological University, Singapore via the Presidential Postdoctoral Fellowship (M.L.), the start-up funding (K.J.H., Grant 002271-00001), NAP Award 020482 (Y.W.) and from The University of Birmingham via the start up funding (M.L.).


\bibliography{reference}

\clearpage
\setcounter{equation}{0}
\renewcommand{\theequation}{A\arabic{equation}}%
\setcounter{figure}{0}
\renewcommand{\thefigure}{A\arabic{figure}}%
\renewcommand{\figurename}{\EDF}%

\onecolumngrid
\begin{appendix}

\setcounter{tocdepth}{2}

\tableofcontents
\newpage
\section{Compression experiment} \label{Chapter_1}

\begin{figure*}[h]
    \centering
    \includegraphics[width = 0.75\linewidth]{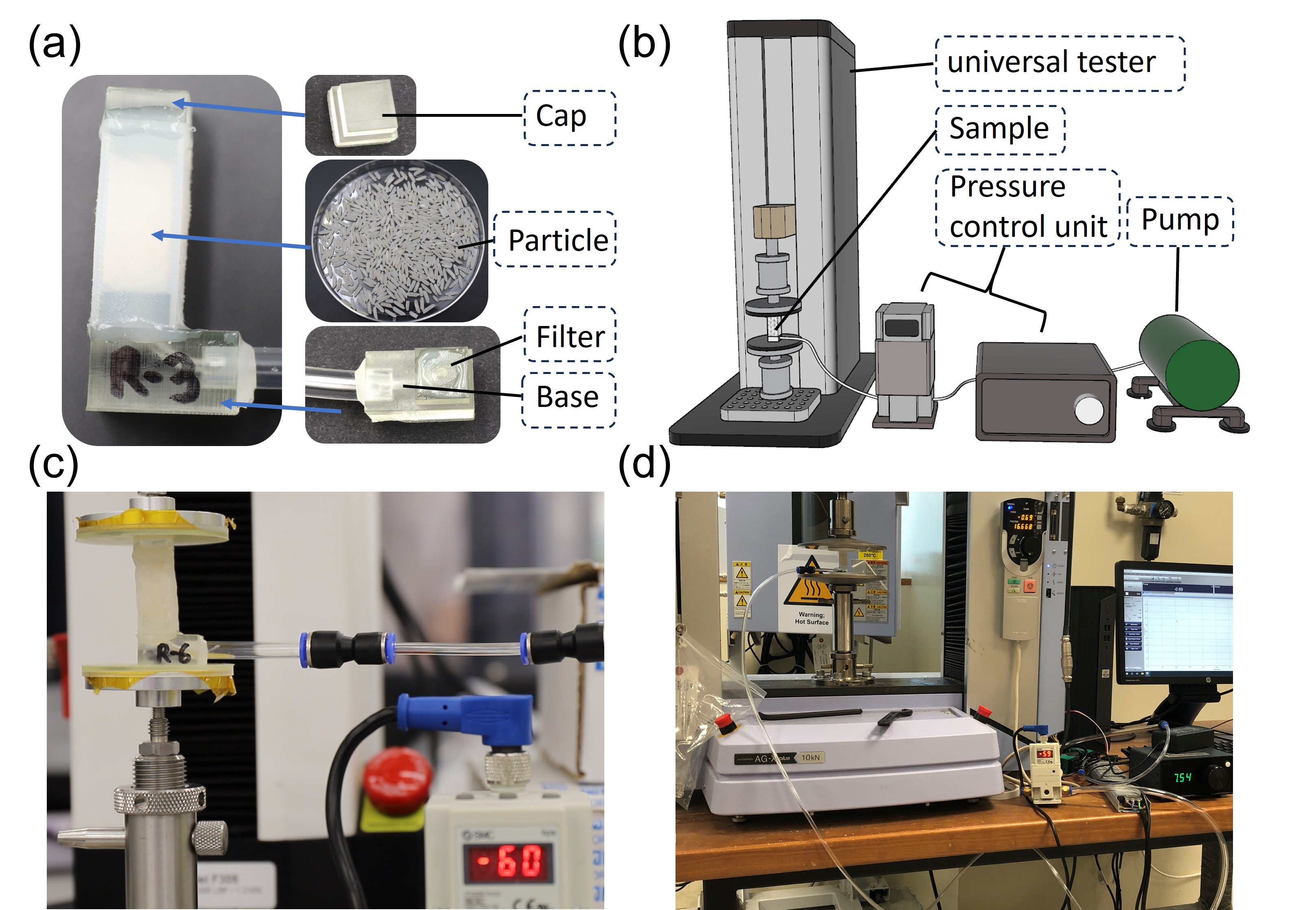}
    \caption{\textbf{Sample and the compression experiment setup} 
    \textbf{(a)} Components of the tested sample. \textbf{(b)} The schematic diagram of the set up used to control the vacuum pressure during the compression test. \textbf{(c)} Compression test set up using the universal tester Mark-10 with 250 N load cell. \textbf{(d)} Compression test set up using the universal tester Shimadzu with 10 kN load cell.}
    \label{SI1_compressionexperiment}
\end{figure*}

In this section, we show our experiment set up and some supplemental results. Figure~\ref{SI1_compressionexperiment}(a) demonstrates the component of a typical 10 mm $\times$ 10 mm $\times$ 20 mm sample. The cap and the base were additive manufactured with the clear resin with a elastic modulus of 2 GPa after healing using the Form lab 3 stereolithography (SLA) printer. The base was designed to have a 1mm diameter air channel to enable vacuum pressure. A filter was attached to the base by glue to prevent particle leakage. The membrane in the middle was manufactured by molding with the silicone material ECO-Flex 00-50 which has a elastic modulus of 84.8 kPa. Both the cap and the bottom was attached to the silicone membrane by silicone glue after the sample was filled with target granular particles. Figure~\ref{SI1_compressionexperiment}(b) is the schematic diagram of the experiment set up by which we performed compression test on samples while maintaining applying a constant vacuum pressure to the sample. The compression experiment is performed by a universal tester: Mark-10 Model F305 with a 250 N load cell, as shown in Fig.~\ref{SI1_compressionexperiment}(c). Two fixtures were manufactured by Rasier 3D fused deposition modeling (FDM) printer and were installed on the upper and lower plate of he tester to prevent the horizontal movement of the sample. The sample was then placed between these two fixtures. The vacuum pressure is applied and maintained to $-60$ kPa by connecting the sample with a SMC ITV2050-312L electronic vacuum regulator and a G4BL2472-2T parallel vacuum pump. The set up ensured that the vacuum pressure was maintained to 60 kPa despite the volume of the sample changed during the compression experiment.

Firstly, the sample was applied a vacuum pressure of $-60$ kPa and turned into jamming state. Then, the sample was placed between two fixtures on the universal tester. A compression test with constant loading velocity of was then performed while the vacuum pressure was maintained to -60 kPa throughout the test. The force and displacement of the sample was recorded by the tester. After the compression test, we applied a positive pressure to the sample to increase the sample size then shake the sample to rearrange the particles in the sample in order to eliminate the effect of loading history. Four different loading velocity, 1 mm/min, 10 mm/min, 100 mm/min, and 1000 mm/min are tested for every sample which are $8.33\times10^{-4}$/s, $8.33\times10^{-3}$/s, $8.33\times10^{-2}$/s, $8.33\times10^{-1}$/s in strain rate to the sample. Three tests were conducted for every loading strain rate and an average value was adopted.

\begin{figure*}[h]
    \centering
    \includegraphics[width=0.7\linewidth]{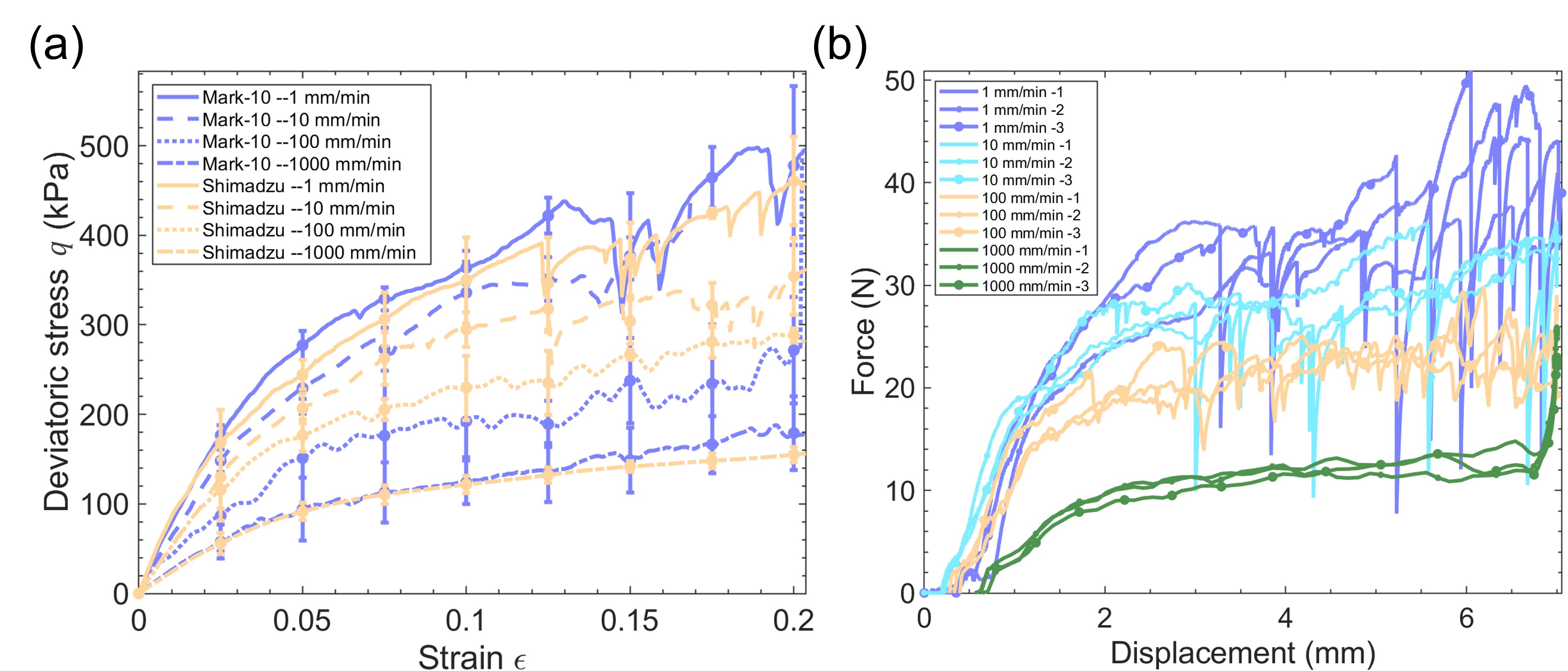}
    \caption{\textbf{Test results with different universal tester.} 
    \textbf{(a)} Comparing the deviatoric stress $q$ vs strain $\epsilon$ of the rice sample tested by Mark-10 and Shimadzu. \textbf{(b)} The untreated data, force-displacement curve, retrieved from the Shimadzu 10 kN load cell for the compression test on the rice sample.}
    \label{SI2_testerresults}
\end{figure*}

When it comes to rate depend mechanical behaviour, the inertia and rigidity of the universal tester may also influence the result. To verify that our results were not influenced by the inertia and rigidity of the universal tester, we performed the compression experiment on the same sample (rice particle) with the same pressure control unit using another large size universal tester: Shimadzu AG-X plus with a 10 kN load cell, as shown in Fig. \ref{SI1_compressionexperiment}(d). As can be seen in the Fig. \ref{SI2_testerresults}(a), which plots the mean value and error bar of the deviatoric stress versus strain under different loading velocity, the results from both two universal tester shows the similar rate weakening behavior. We can conclude that the this rate weakening behavior is not caused by the difference of the testing platform. Additionally, as shown in Fig. \ref{SI2_testerresults}(b) which plots the force-displacement data retrieved directly from the Shimadzu tester, a 0.5 mm space was left between the sample and the tester to reduce the influence of the initial accelerating period of the tester. The loading plate contacts the sample only after it has a displacement of around 1mm. Even though, at the end of the experiment, the inertia of the tester still influenced the results. As can be seen at the green curve (1000 mm/min) in the Fig. \ref{SI2_testerresults}(b), when the tester tried to stop the test at a high loading velocity, the inertia of the tester will cause a significant increase of the force. Therefore, although we compressed the sample to a strain of 0.35, we only looked at the data from the strain of 0 to 0.25.

\section{Additional results from the compression test} \label{chapter_2}   

\begin{figure*}[h]
    \centering
    \includegraphics[width=0.7\linewidth]{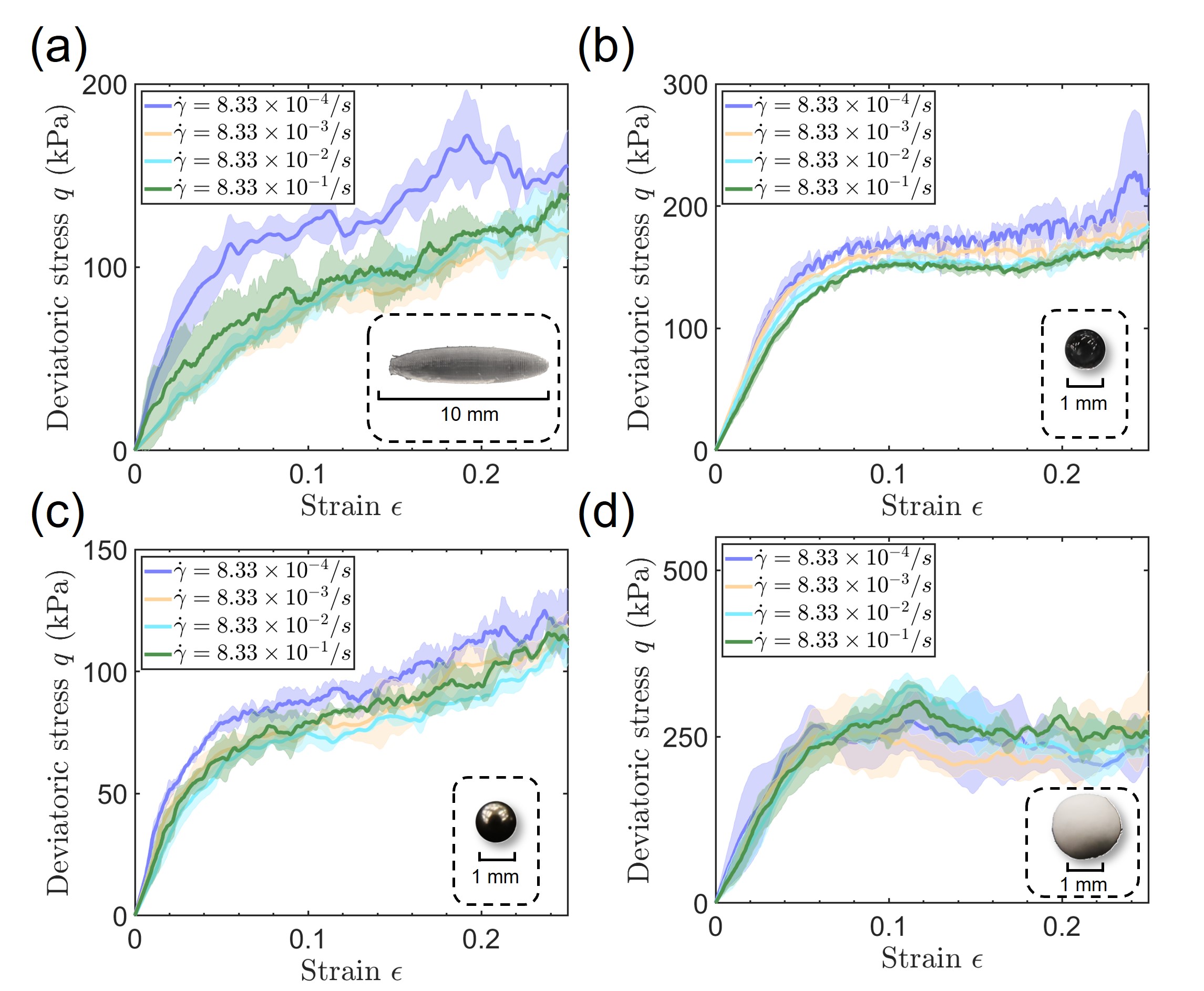}
    \caption{\textbf{The results of other four materials included in the Fig. 2 but not showed in the main text } 
    \textbf{(a)} Deviatoric stress $q$ vs strain $\epsilon$ of the resin printed ellipsoid rice particle with $a = b = 1.8$ mm and $c = 9$ mm. \textbf{(b)} Deviatoric stress $q$ vs strain $\epsilon$ of the 1mm diameter glass beads. \textbf{(c)}  Deviatoric stress $q$ vs strain $\epsilon$ of the 1 mm diameter steel beads. \textbf{(d)} Deviatoric stress $q$ vs strain $\epsilon$ of the sago particles with a mean diameter of 2.05 mm.}
    \label{SI3_otherfourmaterials}
\end{figure*}

In this section, we demonstrate results from the compression test for other materials which can be the supporting information to our research. Figure \ref{SI3_otherfourmaterials} shows the deviatoric stress versus strain for other four material which is included in Figs. \ref{fig:2}(a) and (b) but not shown in the main text. The printed particles are made by clear resin using the Formlab 3 SLA printer. It has a ellipsoid shape with the long semi-axis $a$ equals 9 mm, two short semi-axis $b$ and $c$ equal to 1.8 mm. As can be seen in Fig. \ref{SI3_otherfourmaterials}(a), the resin printed rice also shown a clear rate weakening behavior. It should be noticed that the stick-slip behavior is not obvious for the resin printed rice which agrees with the negative but low magnitude $(A-B)$ value in the RSF law as shown in Fig. \ref{fig:3}(c). Figs. \ref{SI3_otherfourmaterials}(b) and (c) are the $q-\epsilon$ curves for the 1mm diameter glass beads and 1mm diameter steel beads, respectively. Both of them shows a slightly decrease in stress with the increase of the strain rate. However, these stress decrease is too small comparing to the other stress weakening material in this strain rate region so that we treated them as strain independent material in this strain rate region. The sago particles with a average diameter of 2.05 mm were also tested as shown in Fig. \ref{SI3_otherfourmaterials}(d). The sago particle is also time independent material in this loading rate region.

\begin{figure*}[h!]
    \centering
    \includegraphics[width = 0.7\linewidth]{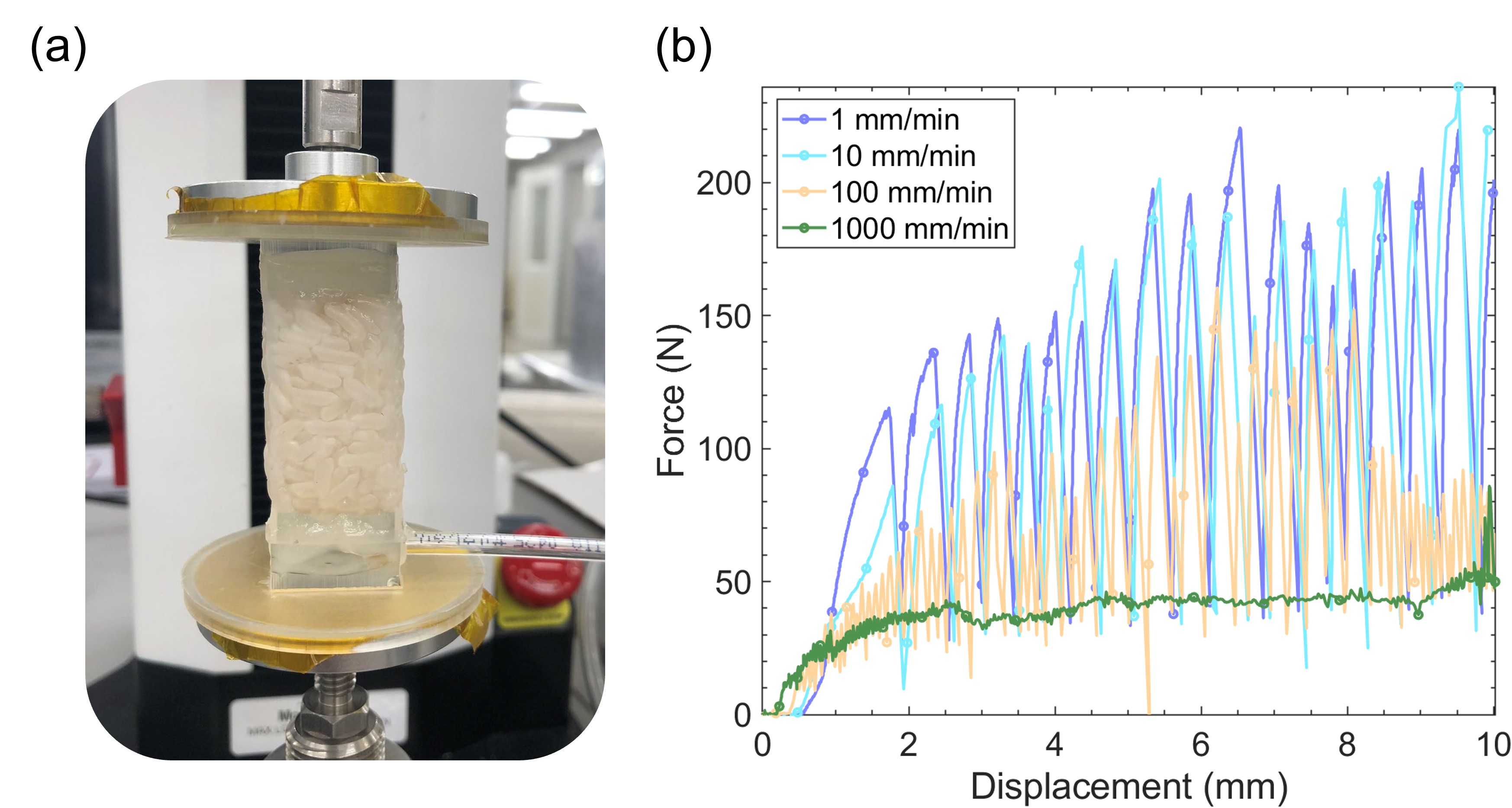}
    \caption{\textbf{Test results for the large rice sample with dimensions 20 mm $\times$ 20 mm $\times$ 40 mm.} 
    \textbf{(a)} The diagram of the vacuumed 20mm$\times$20mm$\times$40mm rice sample under the compression test using the Mark-10 universal tester \textbf{(b)} The untreated data, force-displacement curve, retrieved from the Mark-10 for the compression tests.}
    \label{SI4_largesample}
\end{figure*}

The rice has the largest particle size among our tested granular materials with an average semi-axis length of 7.545 mm, 1.696 mm, and 2.028 mm when we considering it as an ellipsoid shape particle. One may wondering that whether the sample size (the width and the height of the membrane) has influence to this rate dependent behavior. Therefore, we created four larger samples with a 20 mm $\times$ 20 mm $\times$ 40 mm space to fill the granular particles to conduct the same rate dependent compression test, as shown in Fig. \ref{SI4_largesample}(a). The result is plotted in Fig. \ref{SI4_largesample}(b). It can be seen that both the initial elastic modulus and the yield were also decreased with the increase of the loading velocity. However, the stick-slip behavior under low loading velocity has a much larger frequency and magnitude. Comparing with the original 10 mm $\times$ 10 mm $\times$ 20 mm sample, because the sample size increased 2 times in every dimension, the maximum force it can bear also increased from 50.9 N to 232.5 N. We believe that this increase of the stick-slip behavior is caused by this significantly increased force cause the breakage of the rice particles. As shown in {\color{blue}Supporting Video S1}, a large noise can be heard when the slip happened for low velocity loading. Besides, we found many broke particles and large segment for sample after low velocity loading. The increased load level for the large sample also increased the local force between rice particles to exceed its maximum bear level and caused the breakage of particles. To avoid particle breakage, we used the 10 mm $\times$ 10 mm $\times$ 20 mm sample instead of this larger sample for our experiments.

\section{Set up for the measurement of friction coefficient} \label{chapter_4}  

\begin{figure*}[h]
    \centering
    \includegraphics[width=0.95\linewidth]{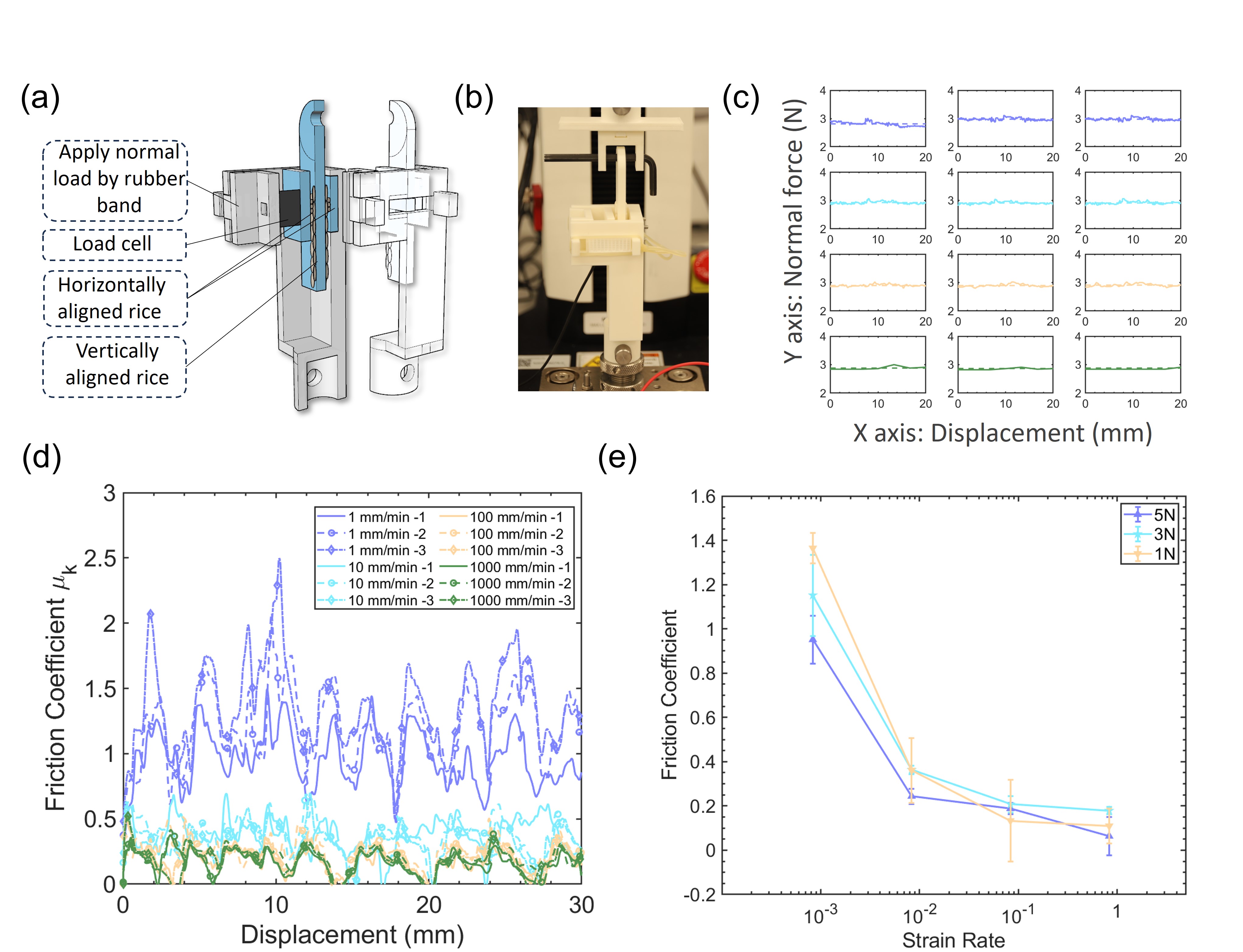}
    \caption{\textbf{The measurement of the friction coefficient  } 
    \textbf{(a)} The schematic diagram of the experiment set up used to measure the friction coefficient $\mu$ of the rice particles under different velocity. \textbf{(b)} The diagram of the experiment set up of the measurement of the friction coefficient $\mu$. \textbf{(c)} The diagram of the applied normal force ($3N$) during the measurement of friction coefficient $\mu$  \textbf{(d)} The plot of the friction coefficient $\mu_k$ under loading velocity of 1 mm/min, 10 mm/min, 100 mm/min and 1000 mm/min.  \textbf{(e)} Comparing the surface friction coefficient of rice particles under normal force of $1N$, $3N$, $5N$.}
    \label{SI5_FrictionCoefficient}
\end{figure*}

An equipment is developed to test the surface friction coefficient of the granular particles. We try to eliminate the effect of the shape locking and dilation while investigating the influence of loading velocity to the surface friction of granular particles so that the rheometer is not used here. Fig. \ref{SI5_FrictionCoefficient}(a) shows the schematic diagram of the set up and Fig. \ref{SI5_FrictionCoefficient}(b) shows the set up during test. We glue the rice particles horizontally and vertically to the side plates and middle plate respectively in order to get a contact surface as flat as possible. The inner side plate is fixed to the frame with screw while the outer side plate can only move horizontally in the designed channel. The middle plate is connected to the universal tester Mark-10 so that we can shear it with a constant velocity and can record shear force with it. A 100N Simbatouch load cell is connected to the outer side plate so that it can measure the normal force during the shearing test. Two rubber bends are used to apply the normal pressure because the elastic modulus of the rubber bands are around 1.2 MPa which is significantly smaller than the frame and the rice particles. Therefore, the horizontal displace during the experiment will not cause the change of the normal load so that we can apply a almost constant normal load during the measurement. As can be seen from the Fig. \ref{SI5_FrictionCoefficient}(c), the osculation of the normal load is less than $5\%$ of the designed normal force level.

As can be seen from the {\color{blue}Supporting Video S2}, after applying a normal load, the middle plate will be sheared upward with a constant shearing velocity. The normal force and the shear force will be recorded simultaneously. By dividing the shear force $F_i$ with the normal force $N_i$ at the time interval $\Delta t_i$, we can get the friction coefficient $\mu^i_k$ at that time interval. Figure \ref{SI5_FrictionCoefficient}(d) plots the friction coefficient $\mu_k$ with the shear distance. It should be noticed that because we sheared the middle plate with a constant velocity, we treated the friction coefficient as a dynamic friction coefficient. It can be seen that their still is a fluctuation happening in the friction coefficient, especially in the low loading velocity region, which is caused by the stick-slip behavior. The representative dynamic friction coefficient is calculated by taking an average value of all the friction coefficient during the sliding from 0mm to 30mm, as shown in Eq. (1) in the main text. Additionally, as shown in Fig. \ref{SI5_FrictionCoefficient}(e), we compare the friciton coefficient of the rice particle under normal load of $1N$, $3N$, and $5N$. The results shows that in this normal load region, the decreasing of friction coefficient with increasing loading velocity is independent with the magnitude of the normal load.

\section{Influence of the particle shape on the rate-dependent behavior } \label{chapter_3}   
\begin{figure*}[h]
    \centering
    \includegraphics[width=0.7\linewidth]{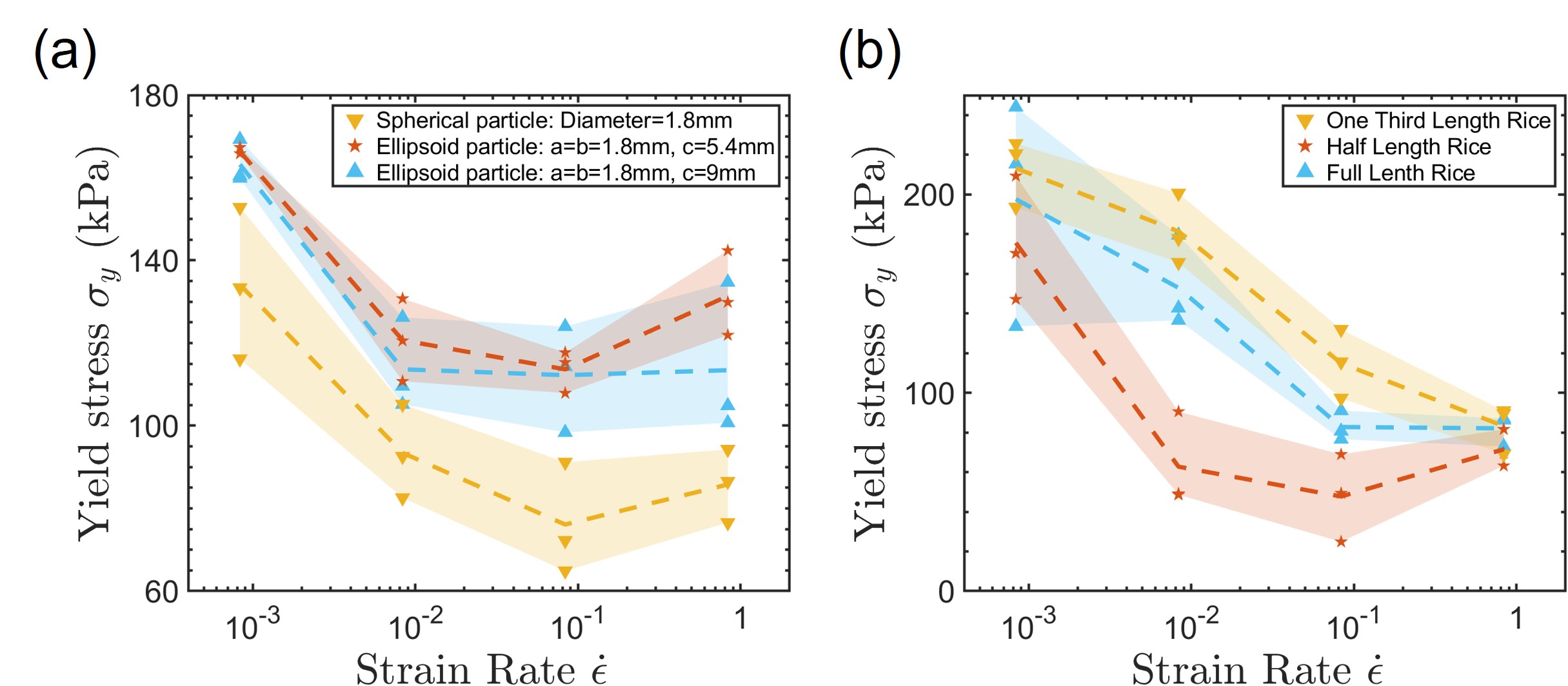}
    \caption{\textbf{The influence of the particle shape to the rate dependent behavior (the used shape is listed on the top of the plot)} 
    \textbf{(a)} the yield stress $\sigma_y$ vs strain rate $\dot{\epsilon}$ for resin particles manufactured by SLA printing. Resin particles has three shapes: Spherical particles whose diameter is 1.8 mm, Ellipsoid particles with $a=b=1.8$mm, $c = 5.4$ mm, and Ellipsoid particles with $a = b = 1.8$ mm, $c = 9$ mm ($a, b, c$ are the three semi-axis of the ellipsoid). \textbf{(f)} the yield stress $\sigma_y$ vs strain rate $\dot{\epsilon}$ of rice particles when they are in full length state, cut in half state, and cut in one third state}
    \label{SI8-shape}
\end{figure*}

Based on the previous research that the particle shape is one of key factors to the shear response of a particle cluster, we conduct experiments to investigate the effect of particle shape on this rate-softening behavior. 
By using SLA additive manufacturing method, we fabricated resin particles with three shapes: ball particles with 1.8 mm diameter, ellipsoid particles with 1.8 mm in both short axis and 5.4 mm in long axis ($1:3$), and ellipsoid particles with 1.8 mm in both short axis and 9 mm in long axis ($1:5$). The particle was printed by using FormLab Form 2 printer with Clear Resin material. It should be mentioned that this three groups of printed particles were tested in 20 mm $\times$ 20 mm $\times$ 40 mm sample in order to reduce the boundary effect due to their larger particle size.
As shown in Fig. \ref{SI8-shape}(a), although the yield stress level is not the same for the three shape particles, they have very similar strain rate softening behavior, especially when the strain rate decrease from $8.33\times{10}^{-4}$/s to $8.33\times{10}^{-3}$/s where the largest drop of yield stress happens. The slop for these ball shape, $1:3$ ellipsoid and $1:5$ ellipsoid are $-25.483$, $-26.348$ and $-29.012$ respectively. From $8.33\times{10}^{-3}$/s to $8.33\times{10}^{-1}$/s, the yield stress meets the lower boundary for all three group and not showing as sharp decrease as the former range. The slightly increase from $8.33\times{10}^{-2}$/s to $8.33\times{10}^{-1}$/s referring to it reaches the boundary of this rate softening behavior which is also witnessed in same strain rate range in the friction coefficient test on resin in Fig. \ref{fig:3}(c). 
Additionally, we also tried to cut the rice particles into half and one third and to reduce the diameter to length ratio and aspect ratio. The shape of the particle is showed on the top of the plot. We made a cutting mold with the shape of rice which is 8mm in long axis. We picked the rice particle that can fitted in the mold and cut them in to half and one third with the help of the mold. The average length of the long axis of these three particles are 7.55 mm, 3.74 mm, and 2.41 mm respectively.  We then conducted compression test on them with a confining pressure of 60 kPa.  The results shown in Fig. \ref{SI8-shape}(b) shows that they also have a similar trend of stress softening behavior while their particle length greatly decreased. Based on these experiment results, it seems that as long as the particles still have low angularity which provides sufficient contacting area, the shape of the particle will not have significant influence on this strain-rate softening behavior.

\section{Set up for the rheometer experiment} \label{chapter_5}   
The rheometer shear experiment setup is as shown in Fig. \ref{fig:4}(a). Considering the stress level of this experiment is not as large as the compression experiment, we used a 20 mm $\times$ 20 mm $\times$ 40 mm sample here. The bottom of the sample was fixed while the top of the sample was glued to the PP25 measuring probe of the Anton Paar MCR 302 Rheometer. To ensure that the rotating axis of the measuring probe aligns with the center of the sample, two 3D-printed connecting parts are glued to both surfaces of the sample. All parts are glued to each other through UV-cross-linked resin. The gap distance between the top measuring plate and the glass bottom substrate was set as $d = 75$ mm so that the rice chamber has a length of 40 mm, as sketched in Fig. \ref{fig:4}(a). The rheometer is not applying any normal pressure to the top plate so that the pressure acting on the top plate is equal to the vacuum pressure. By using the rheometer, the top plate rotates with a fixed angular velocity $\omega = \dot{\theta}$ to apply shear while the resultant torque $T$ was measured simultaneously. The shear stress $\tau$ is converted from the torque $T$ following
\begin{align}
    \tau\equiv\frac{4}{3{\pi}R^3}T,
\end{align}
where $R = 12.5$ mm is the radius of the measuring probe. The shear rate $\dot{\gamma}$ is converted from the angular velocity $\omega$ following
\begin{align}
    \dot{\gamma}=\frac{2R}{3d}\omega,
\end{align}
where $R = 12.5$ mm is the radius of the measuring probe, and $d = 75$ mm is the distance between the top measuring plate and the bottom glass substrate. The shear strain is defined as
\begin{align}
    \gamma=\dot{\gamma}t,
\end{align}
where $t$ is the time of the measurement.

Using the same set up shown in Fig. \ref{SI1_compressionexperiment}(b), the air pressure is maintained to a constant value during the rheometer test. We adopted 4 different vacuum pressure in the experiment including the 0 kPa, 20 kPa, 40 kPa, and 60 kPa. For each pressure, the sample was subjected to 4 different shear rates $10^{-1}$/s, $10^{-2}$/s, $10^{-3}$/s, and $10^{-4}$/s. Before each measurement, the air pressure difference between the rice chamber and the external environment was set as 0, and the rice chamber was squeezed multiple times to reset the configuration of the rice packing. The vacuum air pressure was then set to the value of interest. The shear is then applied by rotating the top probe with a constant angular velocity $\omega$, which relates to $\dot{\gamma}$ as shown in Eq. (A2). Each experiment is repeated three times. To quantify the rate dependence, we examine the secant modulus measured at $\gamma = 10\%$, which is defined as the ratio between $\tau$ and $\gamma$ at $\gamma = 10\%$, as
\begin{align}
    G_s=\frac{\tau_{\gamma = 10\%}}{\gamma = 10\%}.
\end{align}
\section{FEM simulation} \label{chapter_6}   
\begin{figure*}[h]
    \centering
    \includegraphics[width = 0.7\linewidth]{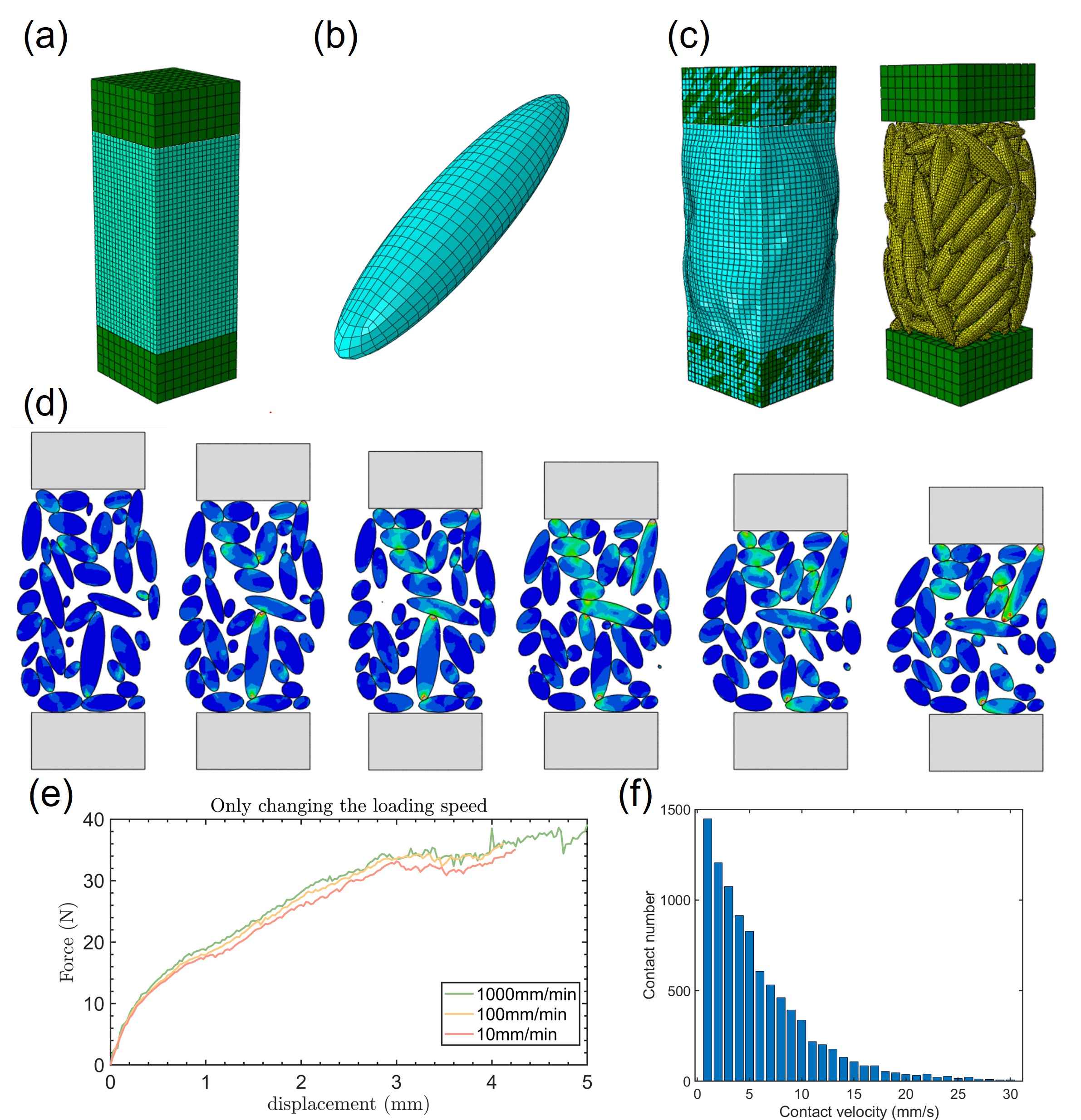}
    \caption{\textbf{The FEM simulation for the compression test } 
    \textbf{(a)} The top and bottom cap of the sample was simulate by rigid body while the membrane was simulated by the shell element. \textbf{(b)} The rice particles was simulated by ellipsoid particles with three semi-axis $a=1.6$ mm, $b=2$ mm, and $c=8$ mm, respectively. \textbf{(c)} (Left) The sample after compaction and was applied 60 kPa vacuum pressure. (Right) The particle assembly inside the sample. \textbf{(d)} The cross section view of the middle Y-Z plane during the compression. \textbf{(e)} Force-Displacement curve for the simulation with three different loading velocity: 1000 mm/min, 100 mm/min, 10 mm/min. The other parameters of these three simulations are the same. \textbf{(f)} The velocity distribution of the contact mesh number during the simulation of loading velocity 1000 mm/min (16.7 mm/s) with friction coefficient $\mu = 0.2$.}
    \label{SI7_FEM}
\end{figure*}

We intend to use FEM simulation to verify our argument on the change of the force chain due  to the rate dependent surface friction cause this rate weakening behavior. The FEM is adopted instead of DEM is that the relative low modulus of rice, which is around 100 MPa by compression test, may cause the particle to deform during the compression which can not be properly simulated by DEM. We are using the Explicit solver in the Abaqus 2020 to conduct this simulation.
To simulate the vacuumed compression test on rice particles, we built two 10 mm $\times$ 10 mm $\times$ 5 mm rigid body plates as the top and bottom 3D printed cap while a 10 mm $\times$ 10 mm $\times$ 20 mm hollow shell element with 1 mm thickness was connected in between to simulate the silicone membrane, as shown in Fig. \ref{SI7_FEM}(a). 135 ellipsoid particles whose three semi-axis are $a = 1.6$ mm, $b = 2$ mm, and $c = 8$ mm are used to simulated the rice particles as shown in Fig. \ref{SI7_FEM}(b). The density and elastic modulus of the membrane are $\rho_m = 1 $ kg/m$^3$, $E_m = 1.5$ MPa while the density and elastic modulus of the particles are $\rho_p = 1.1$ kg/m$^3$, $E_m = 100$ MPa. Hexahedral mesh C3D8R with 1 mm width was used on the top cap and bottom cap to save the computational power. Quadrilateral mesh S4R with 0.5 mm width was used on the membrane. Hexahedral mesh C3D8R with 0.25 mm width was used on the ellipsoid particles.
Before the compression simulation, a compaction procedure was conducted to get a particle assembly that can represent the rice particles in the tested sample. In the first step, we apply a 0.1 N pressure normal to the silicone membrane pointing outside to make the membrane expand. In this step, we do not add any contact relation to the particles and the membrane so that the membrane can pass through the particles freely. Then, in the second step, we reverse the direction applied pressure on the membrane and add a hard contact between the particles and the membrane so that the membrane will move inwards and drive the particles to form a compacted state. The friction coefficient is set to zero during these two steps in favourite to the particle movement. Finally, in the third step, a 60 kPa confining pressure was applied on the normal direction to the membrane to simulate the applied vacuum pressure while the friction coefficient is applied. The compacted model with a confining pressure of 60 kPa, as shown in Fig. \ref{SI7_FEM}(c) was then used for the compression simulation.   
During the compression step, the upper cap is moved downward with a constant velocity. A general contact with a constant friction coefficient $\mu$ is applied to the model as well. We firstly conducted three simulations with same friction coefficient $\mu = 0.2$ but with different loading velocity of 1000 mm/min, 100 mm/min, and 10 mm/min, respectively. As shown in Fig. \ref{SI7_FEM}(e), these three simulations had similar results indicating the effect of inertia can be neglected. Therefore, to save the computational power, we used a constant loading velocity of 1000 mm/min (strain rate $8.33\times{10}^{-1}$/s to the particle media) in the simulation while changing the friction coefficient to represent the different loading rate. Figure \ref{SI7_FEM}(d) shows the middle cross section view of the compression for $\mu=0.2$. {\color{blue}Supporting Video S3} shows the section view of the middle plane of the sample during the compression. It should be mentioned that we also tracked the relative velocity of contacting mesh number with subroutine in the simulation with $\mu=0.2$ and the result is shown in Fig. \ref{SI7_FEM}(f). It can be seen that the local relative velocity of the contacting surface is a distribution and can not be simply represented by the loading velocity of the cap. However, our main purpose here is to qualitatively verify our argument that the weaken of the force chain due to the reduce of the surface friction causes this strain rate weakening behavior. In the future work, we may build a better model to describe the decreasing strength with strain rate quantitative.   

\end{appendix}

\end{document}